\documentclass[%
 reprint,
 superscriptaddress,
 amsmath,amssymb,
 aps,
 prd,
 floatfix,% emergency float placement, avoids floats stuck to the end
]{revtex4-2}

\usepackage{graphicx}% Include figure files
\usepackage{dcolumn}% Align table columns on decimal point
\usepackage{bm}% bold math
\usepackage{booktabs}% nicer table rules
\usepackage{multirow}% multi-row table cells
\usepackage{hyperref}% add hypertext capabilities
\usepackage{color}
\usepackage{float}
\usepackage{soul}
\usepackage{xcolor}
\graphicspath{{Pics/}}% figures are stored in the Pics sub-folder

\begin{document}

\title{Prediction of the Geoneutrino Signal at the Jinping Neutrino Experiment}

\author{Kaikai Meng} %\email[First author: ]{mkkpaper@foxmail.com}。% let's keep one first author and one Corresponding author; PRD does not require the email of the first author %
\author{Yang Zhang} \email[Corresponding author: ]{yangzhangsdu@email.sdu.edu.cn}
\affiliation{Key Laboratory of Particle Physics and Particle Irradiation (MOE), Institute of Frontier and Interdisciplinary Science, Shandong University, Qingdao, Shandong 266237, China}

\date{\today}

\begin{abstract}
We present refined predictions of the geoneutrino flux for the 500-tonne Jinping Neutrino Experiment (JNE). We construct a local crustal model with a resolution of \(0.2^\circ\times0.2^\circ\) based on the SWChinaCVM-2.0 seismic velocity model and integrate it with CRUST1.0 to establish our reference model. For comparison, we also calculate the geoneutrino flux using ECM1 and CRUST1.0 as standalone global crustal models. Assuming an energy scale of 500~p.e./MeV and an exposure of 500~t$\times$1500~days, we predict the inverse beta decay event spectrum, including both the geoneutrino signal and the reactor antineutrino background. The distinct spectral shapes of the $^{238}$U, $^{232}$Th, and reactor antineutrino components allow their respective contributions to be obtained through spectral fitting. These predictions provide inputs for the JNE data analysis, particularly for extracting the mantle geoneutrino signal and testing competing bulk silicate Earth models.
\end{abstract}

\maketitle

%============================================================
\section{Introduction}
%============================================================

The total radiogenic heat production of the Earth remains a fundamental unresolved question in geophysics. Radioactive decay of the heat-producing elements (HPEs)—potassium, thorium, and uranium—contributes to the Earth's internal energy budget and plays a key role in its thermal evolution, with implications for mantle convection, plate tectonics, and the geodynamo~\cite{Dye:2012}. Competing bulk silicate Earth (BSE) models predict a broad range of radiogenic powers, commonly grouped into three categories: low-$Q$ ($11 \pm 2$~TW)~\cite{Javoy:2010}, medium-$Q$ ($20 \pm 4$~TW)~\cite{ArevaloJr:2010}, and high-$Q$ ($33 \pm 3$~TW)~\cite{turcotte:1984}.

The Earth's total surface heat flow is estimated to be $46 \pm 3$~TW~\cite{Jaupart:2007, JH:2010}. This heat loss reflects both ongoing radiogenic heating and the release of stored internal energy, including primordial heat inherited from planetary formation. Because the nonradiogenic contribution remains poorly constrained, surface heat flow alone cannot determine the Earth's radiogenic heat production. 

The Earth consists of the core, mantle, and crust. Models of its thermal evolution generally assume negligible HPE abundances in the core, attributing radiogenic heat production to the crust and mantle. Crustal HPE abundances can be estimated from rock samples and inferred indirectly from seismic data using lithological mixing models and geochemical end-member compositions~\cite{Huang:2013}. Mantle abundances, however, are less directly constrained. The deepest borehole extends only about 12~km below the surface~\cite{Zhu:2026}, well short of the mantle at that location. Geochemical estimates of mantle composition therefore rely largely on mantle-derived rocks brought to the surface by geological processes. These samples provide limited spatial coverage and may not be representative of the mantle as a whole. 

Discriminating among BSE models requires observational constraints on the abundance of HPEs in the deep Earth. Geoneutrinos—predominantly electron antineutrinos, but also electron neutrinos emitted by radioactive decays within the Earth—provide a unique probe of these otherwise inaccessible reservoirs~\cite{McDonough:2014}. The mantle contribution can be inferred by subtracting the predicted lithospheric signal from the measured total geoneutrino signal. Given assumptions about the spatial distribution and relative abundances of HPEs, this residual can be translated into mantle radiogenic heat production and compared with the predictions of competing BSE models. 

Experimental efforts to constrain the Earth's radiogenic heat budget are already underway. KamLAND in Japan and Borexino in Italy have observed geoneutrinos~\cite{KamLAND-2005,KamLAND-2011,KamLAND-2013,KamLAND-2022,Borexino-2010,Borexino-2013,Borexino-2020}. Nevertheless, limited event statistics and backgrounds continue to constrain the measurement precision, with the most precise results reaching relative uncertainties of approximately 15--20\%. The corresponding uncertainties in the measured signal were comparable to the predicted difference between the low- and high-$Q$ models, limiting their ability to discriminate among BSE models. SNO+ has extended geoneutrino measurements to North America, providing an observation in the Western Hemisphere~\cite{SNO+-2025-1, SNO:2025chx}. JUNO in China has also begun operation and data taking, opening new prospects for geoneutrino measurements~\cite{JUNO:2025gmd}. 

The China Jinping Underground Laboratory (CJPL), located 2400~m underground in Sichuan Province, offers favorable conditions for a next-generation geoneutrino experiment. Its substantial overburden strongly suppresses cosmogenic backgrounds, while its distance of more than 950~km from all nuclear power plants reduces the reactor antineutrino background, a major background in geoneutrino measurements. These conditions yield an exceptionally favorable expected signal-to-background ratio compared with the operating and proposed detectors~\cite{JinpingNE:2016}. The 500-tonne Jinping Neutrino Experiment (JNE) is currently under construction at CJPL. Reliable predictions of the lithospheric geoneutrino flux and reactor antineutrino background at the site will provide essential inputs for future JNE data analysis.

Accurate geoneutrino flux predictions at Jinping require detailed knowledge of the surrounding crust. Nearly 50\% of the geoneutrino signal originates within 300~km of the detector~\cite{Sramek:2016}, making the predicted flux particularly sensitive to local crustal structure and HPE abundances. Global crustal models, typically resolved at $1^\circ$ and combined with uniform HPE abundances within each prescribed crustal category, cannot fully represent this spatial heterogeneity. A dedicated near-field model with finer spatial resolution and locally constrained composition is therefore needed.

Previous predictions of the geoneutrino flux at Jinping relied on global crustal models~\cite{Sramek:2016,Wipperfurth2020,Wan:2017}. None of these studies incorporated a dedicated near-field crustal model constrained by regional data, leaving the impact of local refinement on the predicted flux unquantified. To address this gap, we construct a near-field geophysical model of the Jinping region with a spatial resolution of $0.2^\circ\times0.2^\circ$, using seismic constraints from SWChinaCVM-2.0~\cite{Liu2023}. We integrate this model with CRUST1.0~\cite{CRUST:2014} to obtain our reference geoneutrino flux prediction. Comparisons with predictions based on CRUST1.0 and ECM1~\cite{ECM1:2023} alone allow us to assess the effects of local refinement and global model selection. We also predict the inverse beta decay (IBD) event spectrum for the 500-tonne JNE, including both the geoneutrino signal and the reactor antineutrino background. 

The remainder of this paper is organized as follows. Section~\ref{sec:fluxcalc} outlines the geoneutrino flux calculation. Section~\ref{sec:earthmodel} describes the geophysical and geochemical models adopted in this study. Section~\ref{sec:reactorbkg} details the reactor antineutrino background calculation. Section~\ref{sec:results} presents the results, and Section~\ref{sec:summary} summarizes our findings. 

%\textcolor{red}{}

%============================================================
\section{Geoneutrino Flux Calculation}
\label{sec:fluxcalc}
%============================================================

Geoneutrinos are detected through the IBD reaction on free protons in hydrogen-rich targets, with an energy threshold of $E_{\bar\nu_e}^{\mathrm{thr}} = 1.806$~MeV. The signal is conventionally expressed in terrestrial neutrino units (TNU), where 1~TNU corresponds to one event per $10^{32}$ free protons per year at 100\% detection efficiency. The differential IBD interaction rate per target proton with respect to the electron antineutrino energy ($E$) is
\begin{equation}
    R(E) = \phi(E) \times \sigma_{\mathrm{IBD}}(E),
\end{equation}
where $\phi(E)$ is the differential electron antineutrino flux, and $\sigma_{\mathrm{IBD}}(E)$ is the IBD cross section~\cite{Vogel:1999}. Integrating $R(E)$ above the IBD threshold and multiplying by $10^{32}$ target protons and one year gives the numerical signal value in TNU. %The slow liquid scintillator considered for JNE contains approximately $7.2 \times 10^{31}$ free protons per kilotonne~\cite{JinpingNE:2016}.

The differential geoneutrino flux $\phi_i(E)$ at the detector site from HPE $i$ is calculated by integrating the contributions from all volume elements within the Earth~\cite{Sramek:2016},
\begin{equation}
    \phi_i(E) = \frac{\lambda_i N_A}{\mu_i}\, n_\nu(i)
    \int \frac{A_i(\vec r)\,\rho(\vec r)}{4\pi L^2}\,
    P_{ee}(E,L)\, f_i(E)\; d\vec r,
    \label{Eq:flux_diff}
\end{equation}
where $\lambda_i$ is the decay constant, $\mu_i$ is the standard molar mass, and $n_\nu(i)$ is the number of electron antineutrinos emitted per complete decay chain of HPE $i$, while $N_A$ is Avogadro's constant. The HPE abundance and rock density at each source position $\vec r$ are denoted by $A_i(\vec r)$ and $\rho(\vec r)$, respectively, with the source located at a distance $L$ from the detector. The electron antineutrino survival probability, $P_{ee}(E,L)$, describes the probability that an electron antineutrino with energy $E$ remains in the electron flavor after propagating a distance $L$ in vacuum. In the three-flavor neutrino mixing framework, it is given by~\cite{PDG:2025},
\begin{equation}
\begin{split}
    P_{ee}(E,L) = 1 &- \cos^4\theta_{13}\,\sin^2 2\theta_{12}\,
    \sin^2\!\left(\frac{\Delta m^2_{21}L}{4E}\right) \\
    &- \sin^2 2\theta_{13}\,
    \sin^2\!\left(\frac{\Delta m^2_{31}L}{4E}\right), 
\end{split}
    \label{Eq:Pee}
\end{equation}
where $\theta_{12}$ and $\theta_{13}$ are the neutrino mixing angles, and $\Delta m^2_{ij} = m_i^2 - m_j^2$ is the mass-squared difference. Wan \textit{et al.}~\cite{Wan:2017} found that the uncertainty in $\theta_{12}$ dominates the oscillation-parameter contribution to the uncertainty in the predicted geoneutrino signal. We therefore adopt $\sin^2\theta_{12}=0.3092\pm0.0087$ from the latest JUNO measurement~\cite{JUNO:2025gmd}, while taking the remaining oscillation parameters from the PDG~\cite{PDG:2025}. The same study found that matter effects modify the predicted signal by only about $0.3\%$, so we neglect them here. 

The quantity $f_i(E)$ in Eq.~\ref{Eq:flux_diff} denotes the normalized electron antineutrino energy spectrum per complete decay chain of HPE $i$. Geoneutrinos originate primarily from the uranium and thorium decay chains and the decay of potassium-40~\cite{Enomoto:2005}. Among them, only $^{238}$U and $^{232}$Th produce $\bar{\nu}_e$ above the IBD threshold and therefore contribute to the detectable geoneutrino signal. In contrast, the $\bar{\nu}_e$ spectra from $^{235}$U and $^{40}$K lie entirely below the IBD threshold and thus do not contribute to the detectable IBD signal. The corresponding geoneutrino energy spectra can be found in Ref.~\cite{Wan:2017}. 
%\begin{align*}
%& ^{238}\mathrm{U} \rightarrow ^{206}\mathrm{Pb} + 8\alpha + 6e^- + 6\bar\nu_e + 51.7~\mathrm{MeV},\\
%& ^{235}\mathrm{U} \rightarrow ^{207}\mathrm{Pb} + 7\alpha + 4e^- + 4\bar\nu_e + 46.4\ \mathrm{MeV} ,\\
%& ^{232}\mathrm{Th} \rightarrow ^{208}\mathrm{Pb} + 6\alpha + 4e^- + 4\bar\nu_e + 42.7~\mathrm{MeV},\\
%& ^{40}\mathrm{K} \rightarrow ^{40}\mathrm{Ca} + e^- + \bar\nu_e + 1.31~\mathrm{MeV} \quad(89.3\%),\\
%& ^{40}\mathrm{K} + e^- \rightarrow ^{40}\mathrm{Ar} + \nu_e + 1.51~\mathrm{MeV} \quad(10.7\%).
%\end{align*}

%\begin{figure}
%    \centering
%    \includegraphics[width=0.8\linewidth]{HPEs_energy_spectra.png}
%    \caption{The geoneutrino energy spectra produced by HPEs. The $\bar\nu_{e}$'s are produced in the decays of $^{238}$U, $^{232}$Th, $^{40}$K, and $^{235}$U. $\nu_e$ from $^{40}$K electron capture is not shown. \textcolor{red}{If the plot is given by Enomoto, we could remove it. }}
%    \label{fig:nue_intensity}
%\end{figure}

%============================================================
\section{Earth Model}
\label{sec:earthmodel}
%============================================================

\subsection{Geophysical Models}
\label{sec:earth_models}
The choice of geophysical model involves the selection of both mantle and crustal models.

For the mantle, four scenarios can be considered: (1) a uniform mantle with a homogeneous composition; (2) a layered uniform mantle model comprising a depleted upper mantle and an enriched lower-mantle reservoir, each with a homogeneous composition; (3) a single enriched-blob model, in which the large low-shear-velocity provinces (LLSVPs) are represented by a single enriched reservoir; and (4) a symmetric two-blob model, with two enriched reservoirs beneath the Pacific and Africa. The predicted mantle geoneutrino signals vary by only about $10\%$ among these scenarios~\cite{Sramek:2013}, indicating that the predicted signal is relatively insensitive to the assumed mantle distribution. Given this modest model dependence, we adopt the second scenario, the layered uniform mantle model, as a representative baseline configuration.

For the crust, several global crustal models are available, including CRUST1.0, LITHO1.0~\cite{LITHO1.0-2014}, and the recently developed ECM1 model, which incorporates nearly 20,000 seismic measurements and an updated sediment model. Wipperfurth \textit{et al.}~\cite{Wipperfurth2020} showed that the predicted geoneutrino signal is relatively insensitive to the choice between CRUST1.0 and LITHO1.0, and emphasized that improving the near-field crustal model is more important for reducing the uncertainty in the predicted signal. In the present study, we perform the global crustal calculation using both CRUST1.0 and the more recent ECM1 model, allowing us to evaluate how updates in the global crustal structure affect the predicted geoneutrino signal.

\begin{figure}
	\centering
	\includegraphics[width=0.8\linewidth]{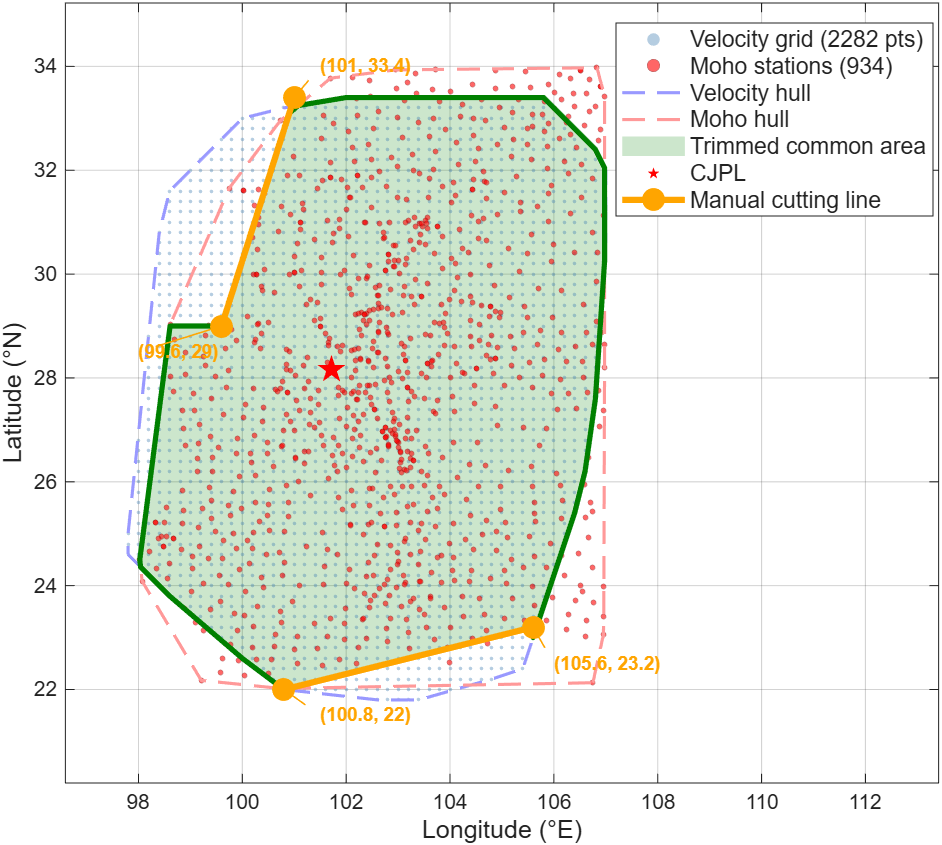}
	\caption{Geographical coverage of the SWChinaCVM-2.0 (blue) and SWChinaCVM-MOHO-2.0 (red) models and the trimmed overlap region (light green) used in this study. Blue and red dots represent the velocity grid points and Moho stations of the respective models. Dashed curves in matching colors outline their convex hulls. The shaded region is the intersection of the two hulls, with two corners removed along the thick orange lines with circular endpoint markers. The red star marks the location of CJPL.}
	\label{SWChinaCVM}
\end{figure}

The high-resolution regional velocity model SWChinaCVM-2.0~\cite{Liu2023} covers the crust and uppermost mantle of the Sichuan--Yunnan region in southwestern China. It was constructed through a joint inversion of approximately 7.06 million body-wave travel-time measurements and 0.1 million Rayleigh-wave phase-velocity dispersion measurements from permanent seismic stations and dense temporary arrays, accounting for topographic effects. SWChinaCVM-2.0 provides three-dimensional $P-$ and $S-$wave velocity ($V_p$ and $V_s$) structures of the crust and uppermost mantle with a horizontal grid spacing of $0.2^\circ \times 0.2^\circ$, substantially finer than the $1^\circ \times 1^\circ$ grids of the global CRUST1.0 and LITHO1.0 models. Its spatial coverage includes the JNE site, as shown in Fig.~\ref{SWChinaCVM}.

The regional Moho interface model SWChinaCVM-MOHO-2.0 was subsequently developed using three new-generation regional velocity models, SWChinaCVM-2.0, CSES-VM1.0~\cite{Wu2024}, and USTClitho2.0~\cite{Han2022}, together with receiver-function analysis~\cite{Liu2025}. Given the regional coverage and $0.2^\circ$ grid spacing of SWChinaCVM-2.0, we combine it with SWChinaCVM-MOHO-2.0 to refine the near-field crustal density structure and the HPE abundances of the middle and lower crust for the geoneutrino flux calculation at Jinping. 

\subsection{Integration with the Near-Field Model}

A combined model, hereafter referred to as the Combined Model, is constructed for geoneutrino flux calculations at the JNE site by integrating CRUST1.0, SWChinaCVM-2.0, and SWChinaCVM-MOHO-2.0, as described below. 

Fig.~\ref{SWChinaCVM} shows the SWChinaCVM-2.0 velocity grid (blue dots), the Moho stations (red dots), and their respective coverage boundaries (dashed curves). The manually trimmed overlap region (light-green shading) is obtained by intersecting these coverage areas and excluding two corners where sparse Moho-station coverage could compromise the reliability of subsequent Moho-depth interpolation. The trimming boundaries are a southeastern line connecting (100.8$^{\circ}$E, 22$^{\circ}$N) and (105.6$^{\circ}$E, 23.2$^{\circ}$N), and a northwestern line connecting (99.6$^{\circ}$E, 29$^{\circ}$N) and (101$^{\circ}$E, 33.4$^{\circ}$N), both shown as thick orange lines with circular endpoint markers. The resulting overlap region spans 79.19 square degrees and contains 1,957 of the 2,282 velocity grid points and 842 of the 934 Moho stations. 

\begin{figure}
    \centering
    \includegraphics[width=0.85\linewidth]{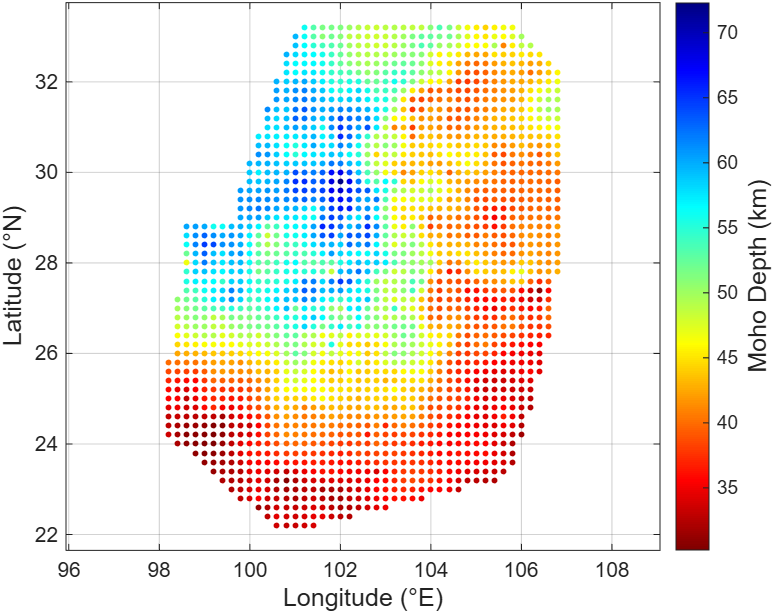}
    \caption{Moho depth map of the near-field region obtained by kriging interpolation onto a $0.2^\circ \times 0.2^\circ$ grid.}
    \label{moho_depth_map}
\end{figure}

First, the Moho depth is determined for each $0.2^\circ \times 0.2^\circ$ grid cell. Because SWChinaCVM-2.0 and SWChinaCVM-MOHO-2.0 differ in spatial coverage, we identify their overlap region and use ordinary kriging~\cite{Deutsch:1998} to interpolate the Moho depths onto the longitude--latitude grid of the velocity model, as shown in Fig.~\ref{moho_depth_map}. All available Moho stations, including those outside the overlap region, are used to fit the semivariogram model $\gamma(h)$ and determine the kriging weights $w_i$ for each target point. Predictions are evaluated only at velocity grid points within the trimmed overlap region. The estimated Moho depth $\hat{Z}(\mathbf{x}_0)$ at a target point $\mathbf{x}_0$ is expressed as a weighted linear combination of the depths $Z(\mathbf{x}_i)$ measured at the $N$ stations, 
\begin{equation}
\hat{Z}(\mathbf{x}_0)=\sum_{i=1}^{N}w_i\,Z(\mathbf{x}_i), 
\end{equation}
where the weights $w_i$ are determined by minimizing the estimation variance subject to the unbiasedness constraint $\sum_{j=1}^{N}w_j=1$. This yields the ordinary kriging equations, 
\begin{equation}
 \sum_{j=1}^{N}w_j\,\gamma(\mathbf{x}_i-\mathbf{x}_j)+\mu=\gamma(\mathbf{x}_i-\mathbf{x}_0), 
\end{equation}
 where $\mu$ is a Lagrange multiplier and $\gamma$ is the semivariogram of the Moho depth field. We adopt an isotropic spherical semivariogram model, 
\begin{equation}
\gamma(h)=\begin{cases}c_0+c\left[\frac{3}{2}\left(\frac{h}{a}\right)-\frac{1}{2}\left(\frac{h}{a}\right)^{3}\right], & 0<h\le a,\\ c_0+c, & h>a,\end{cases}
\end{equation}
where $h$ is the distance between stations, and the nugget $c_0$, partial sill $c$, and range $a$ are determined by fitting the model to the experimental semivariogram of the Moho depth data. 

Second, surface elevations are obtained from the ETOPO1 global relief model~\cite{Amante2009}, which also provides the topographic input for CRUST1.0. This is necessary because SWChinaCVM-2.0 specifies a uniform maximum elevation of 5~km at all grid points rather than the actual surface topography. For each $0.2^\circ \times 0.2^\circ$ grid cell, we average the elevations at nine sampling points from the 1-arc-minute ETOPO1 Ice Surface dataset, following the averaging scheme used in CRUST1.0.

Third, the sediment thicknesses and the relative thickness proportions of the upper, middle, and lower crystalline crust (UC, MC, and LC) are inherited from the parent $1^\circ \times 1^\circ$ CRUST1.0 cell containing each $0.2^\circ \times 0.2^\circ$ grid point. The total crystalline crustal thickness is calculated from the surface elevation, Moho depth, and sediment thickness, and then divided into UC, MC, and LC layers according to these proportions. 

Fourth, representative $V_p$ and $V_s$ values for each layer are calculated as weighted averages of the SWChinaCVM-2.0 velocity samples within its depth range, with weights given by the thickness intervals represented by the samples. For layers containing no sampling points within their depth range, such as thin sedimentary layers, velocities are estimated at the layer midpoint by linear interpolation between the nearest sampling points above and below it. The resulting velocity profile at the JNE site is summarized in Table~\ref{tab:JNE_velocity}. The $V_p$ values increase from approximately 6.0\,km/s in the upper crust to 7.0\,km/s in the lower crust, consistent with typical continental crustal velocity gradients and compositional stratification from a felsic upper crust to a more mafic lower crust.
% The sedimentary layers in the Chengdu basin, northeast of the Jinping site, exhibit substantially lower $V_p$ values of 4.9--5.4\,km/s, compared with 5.4\,km/s at the Jinping site.\textcolor{red}{Chengdu basin vs Jinping, relationship? }. 
%This lateral velocity heterogeneity is captured on the $0.2^\circ$ grid but would be smoothed at the coarser $1^\circ$ resolution of CRUST1.0.  %lateral means same depth. 
\begin{table}[htbp]
\centering
\caption{Seismic velocity profile at the JNE site. }
\label{tab:JNE_velocity}
\begin{tabular}{lcccc}
\toprule
Layer & Thickness (km) & $V_p$ (km/s) & $V_s$ (km/s) & $V_p/V_s$ \\
\midrule
Sediment    & 0.01  & 5.41 & 3.05 & 1.78 \\
UC    & 26.6  & 5.99 & 3.51 & 1.71 \\
MC    & 23.6  & 6.24 & 3.54 & 1.76 \\
LC    & 10.3  & 7.01 & 3.89 & 1.80 \\
\bottomrule
\end{tabular}
\end{table}

Fifth, the density ($\rho$) of each layer is estimated from its representative $V_p$ using the empirical velocity--density relation of Brocher~\cite{Brocher:2005}:
\begin{equation}
\begin{split}
    \rho = 1.6612\,V_p - 0.4721\,V_p^2 + 0.0671\,V_p^3 \\
    - 0.0043\,V_p^4 + 0.000106\,V_p^5,
\end{split}
\end{equation}
where $V_p$ and $\rho$ are expressed in km/s and g/cm$^3$, respectively. 

Finally, the near-field model is integrated into the global CRUST1.0 model. The 93 $1^\circ \times 1^\circ$ CRUST1.0 cells in the near-field region are each subdivided into 25 subcells of $0.2^\circ \times 0.2^\circ$, giving 2,325 subcells in total. These subcells are assigned near-field model parameters where available and otherwise retain all parameters of the parent CRUST1.0 cell. The Combined Model comprises 64,707 CRUST1.0 cells and 2,325 subcells, providing a global crustal model with a refined $0.2^\circ \times 0.2^\circ$ grid in the near-field region for geoneutrino flux calculations. For clarity, the high-resolution crustal model comprising the 1,957 $0.2^\circ \times 0.2^\circ$ grid cells within the trimmed near-field region is hereafter referred to as the Near-Field Model. Its integration with CRUST1.0 yields the previously defined Combined Model. 

\subsection{Geochemical Model}
\label{sec:geochem}
The geochemical model is constructed following Ref.~\cite{Wipperfurth2020}, with HPE abundances assigned separately to each reservoir. Upper crustal abundances are taken from the compilation of Rudnick \& Gao~\cite{RudnickGao:2014}, while middle and lower crustal abundances are inferred from the $V_p$--composition relationships for amphibolite and granulite, respectively (see Eqs.~\eqref{eq:Vp}--\eqref{eq:abundance}). Sedimentary abundances follow the GLOSSII model~\cite{Plank:2014}, oceanic crustal abundances are taken from the global averages of White \& Klein~\cite{WhiteKlein:2014}, and lithospheric mantle abundances are based on the xenolith data of Huang \textit{et al.}~\cite{Huang:2013}. The mantle consists of depleted mantle (DM) and enriched mantle (EM), with EM accounting for 19\% of its mass, following the layered uniform mantle model described in Sec.~\ref{sec:earth_models}. To maintain consistency with this configuration, Monte Carlo simulations in which the U abundance in EM is lower than that in DM are discarded, as the enriched reservoir is expected to carry the higher U abundance. The BSE composition is guided by the U abundance of McDonough \& Sun~\cite{McDonoughSun:1995}, with the U abundance set to 20.3~ppb in this study and the Th and K abundances inferred from the adopted Th/U and K/U ratios~\cite{Arevalo2009,Wipperfurth2018}, respectively. The DM is assigned a U abundance adopted from the compilation of Arevalo \textit{et al.}~\cite{Arevalo2013}, with Th and K abundances derived using the adopted Th/U and K/U ratios inferred from mid-ocean ridge basalt measurements~\cite{Arevalo2009,Wipperfurth2018}. It extends from the base of the lithosphere to 750~km above the core--mantle boundary, beneath which lies the EM reservoir. The EM abundances are determined by mass balance: the lithospheric and DM inventories of each HPE are subtracted from the corresponding BSE inventory, and the remainder is divided by the EM mass. The adopted BSE composition corresponds to a medium-$Q$ model. 

Huang \textit{et al.}~\cite{Huang:2013} found that $V_p$ decreases with increasing SiO$_2$ content in amphibolite- and granulite-facies rocks. Following their approach, we model the middle and lower crust as binary mixtures of felsic (silica-rich) and mafic (silica-poor) end-members, with seismic velocities and HPE abundances expressed as weighted averages of the corresponding end-member values, 
\begin{equation}
V_{\mathrm{model}} = V_f f + V_m m
\label{eq:Vp}
\end{equation}
\begin{equation}
1 = f + m
\label{eq:mass}
\end{equation}
\begin{equation}
a_{\mathrm{output}} = a_f f + a_m m
\label{eq:abundance}
\end{equation}
where $V_f$ and $V_m$ are the $P$-wave velocities of the felsic and mafic end-members, respectively, and $a_f$ and $a_m$ are their abundances of a given HPE. The end-member properties, including $P$-wave velocities and HPE abundances, are taken from amphibolite-facies rocks for the middle crust and granulite-facies rocks for the lower crust. Here, $V_{\mathrm{model}}$ is the $P$-wave velocity provided by the geophysical model, $f$ and $m$ are the felsic and mafic mass fractions, respectively, and $a_{\mathrm{output}}$ is the inferred HPE abundance of the mixture. Because HPE abundances in the middle and lower crust are inferred from $V_p$, adopting SWChinaCVM-2.0 affects the estimated abundances in these layers. 

%============================================================
\section{Reactor Antineutrino Background Modeling}
\label{sec:reactorbkg}
\begin{figure}
    \centering
    \includegraphics[width=0.8\linewidth]{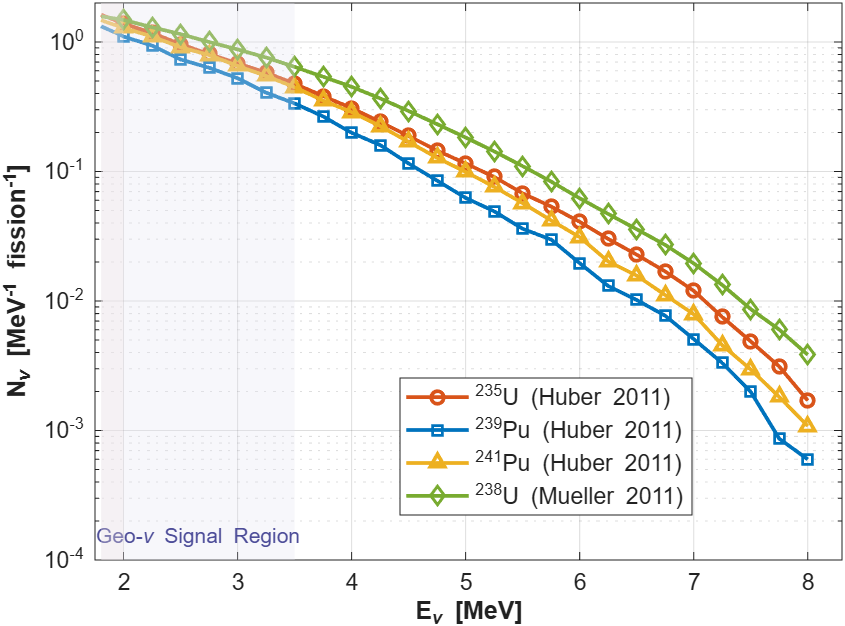}
    \caption{Reactor antineutrino energy spectra for the four principal fissioning isotopes. The spectra of $^{235}$U, $^{239}$Pu, and $^{241}$Pu are taken from Ref.~\cite{Huber:2011}, while that of $^{238}$U is taken from Ref.~\cite{Mueller:2011}.}
    \label{fig:reactor_spectrum}
\end{figure}

%============================================================
Reactor electron antineutrinos constitute a dominant and irreducible background to geoneutrino detection because they produce the same IBD signature. The differential $\bar{\nu}_e$ emission rate ($Q_r(E)$) from a reactor is estimated as 
\begin{equation}
    Q_r(E) = \frac{W_{\mathrm{th}} \cdot LF}{\sum_i f_i e_i}\,
    \sum_i f_i\, S_i(E),
    \label{Eq:reactor_flux}
\end{equation}
where the subscript $r$ denotes the reactor source, $W_{\mathrm{th}}$ is the thermal power, $LF$ is the load factor (set to 1.0 in this study), $f_i$ is the fission fraction of isotope $i$ (with $\sum_i f_i = 1$), and $e_i$ is the energy released per fission of isotope $i$. $S_i(E)$ denotes the antineutrino energy spectrum per fission of isotope $i$, obtained from Refs.~\cite{Huber:2011, Mueller:2011} and shown in Fig.~\ref{fig:reactor_spectrum}. 
Although the IBD threshold is 1.806 MeV, the tabulated spectra in Refs.~\cite{Huber:2011, Mueller:2011} start at 2.0 MeV. To cover the interval between the threshold and 2.0 MeV, we extrapolate the exponential-polynomial parameterizations provided in these references, sampling the spectra at 0.05 MeV intervals. At energies of 2.0 MeV and above, we use the original tabulated data with a bin spacing of 0.25 MeV.

The adopted fission fractions and energies released per fission are taken from Refs.~\cite{Baldoncini:2015} and~\cite{Ma:2012}, respectively, and listed in Table~\ref{tab:EnergyReleased}. Reactor characteristics are obtained from the database maintained by the Istituto Nazionale di Fisica Nucleare (INFN)~\cite{INFN:antineutrino}, which compiles operational data published annually by the International Atomic Energy Agency (IAEA) through its Power Reactor Information System (PRIS)~\cite{IAEA:2015}.
\begin{table}[h]
    \centering
    \caption{Fission fractions $f_i$ and average energies released per fission $e_i$ for the four principal fissioning isotopes.}
    \label{tab:EnergyReleased}
    \renewcommand{\arraystretch}{1.2}
    \begin{tabular}{ccc} \hline\hline
        Isotope & $f_i$ & $e_i$ [MeV/fission] \\\hline
        ${}^{235}$U & 0.58 & $202.36 \pm 0.26$ \\
        ${}^{238}$U & 0.07 & $205.99 \pm 0.52$ \\
        ${}^{239}$Pu & 0.30 & $211.12 \pm 0.34$ \\
        ${}^{241}$Pu & 0.05 & $214.26 \pm 0.33$ \\\hline\hline
    \end{tabular}
\end{table}

%============================================================
\section{Results}
\label{sec:results}
%============================================================

\subsection{Geoneutrino Flux at Jinping}
\label{sec:results_flux}

This section presents the predicted geoneutrino flux at the JNE site. Results for the Combined Model, ECM1, and CRUST1.0 models, each based on \(10^{4}\) Monte Carlo samples, are listed in Tables~\ref{tab:flux_jinping_crust1_0_swchinacvm}--\ref{tab:flux_jinping_crust1_0}, with a breakdown of the contributions from individual crustal and mantle reservoirs shown separately. Central values are the arithmetic mean and uncertainties denote the 68\% percentile interval. 

\begin{figure}[htbp]
	\centering
	\includegraphics[width=1\linewidth]{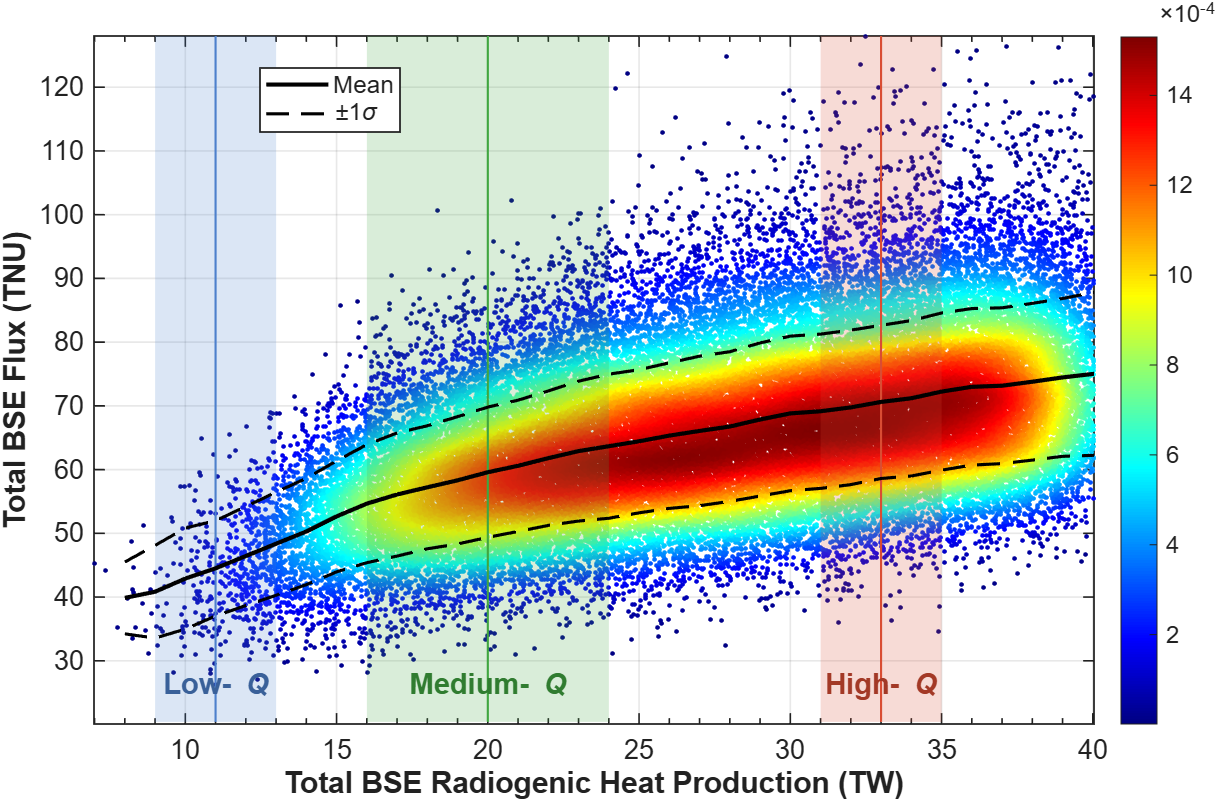}
	\caption{Monte Carlo predictions of the total BSE geoneutrino flux at JNE as a function of total BSE radiogenic heat production, calculated using the Combined Model. The solid black line and the dashed lines denote the mean predicted signal and its $\pm1\sigma$ bounds, respectively, evaluated in sliding windows of 2~TW width with a step of 0.5~TW.}
	\label{fig:mc_signal}
\end{figure}

All Monte Carlo results are based on simulations in which the U abundance in EM exceeds that in DM and the EM abundances of U, Th, and K are all positive. Simulations are excluded if EM is not enriched in U relative to DM or if the combined lithospheric and DM inventory exhausts or exceeds the BSE budget for any of these elements. The retained samples comprise 6,695 of 10,000 simulations (67\%) for the Combined Model, 6,798 of 10,000 (68\%) for ECM1, and 6,694 of 10,000 (67\%) for CRUST1.0. For the $7\times10^{4}$-simulation run used in Fig.~\ref{fig:mc_signal}, 45,134 simulations (64\%) are retained.

\begin{table}[t]
	\centering
	\caption{Geoneutrino flux at the Jinping detector predicted with the Combined Model. }
	\label{tab:flux_jinping_crust1_0_swchinacvm}
	\renewcommand{\arraystretch}{1.4}
	\begin{tabular}{lccc}
		\toprule
		& \multicolumn{3}{c}{Flux (TNU)} \\
		& $U$ & $Th$ & $U + Th$ \\
		\midrule
		Sediment & $0.74^{+0.10}_{-0.10}$ & $0.24^{+0.03}_{-0.03}$ & $0.98^{+0.13}_{-0.13}$ \\
		Upper Crust & $28.39^{+7.24}_{-7.35}$ & $7.38^{+1.18}_{-1.19}$ & $35.77^{+8.36}_{-8.47}$ \\
		Middle Crust & $7.47^{+3.93}_{-3.85}$ & $3.10^{+1.92}_{-1.86}$ & $10.57^{+5.82}_{-5.70}$ \\
		Lower Crust & $1.11^{+0.68}_{-0.69}$ & $0.72^{+0.52}_{-0.57}$ & $1.83^{+1.20}_{-1.26}$ \\
		\hline 
		Crust & $37.72^{+8.65}_{-8.67}$ & $11.44^{+2.51}_{-2.51}$ & $49.16^{+10.89}_{-10.85}$ \\
		Lithospheric Mantle & $1.67^{+1.31}_{-1.36}$ & $0.55^{+0.45}_{-0.47}$ & $2.23^{+1.75}_{-1.83}$ \\
		\hline 
		Lithosphere & $39.40^{+9.20}_{-9.15}$ & $11.99^{+2.72}_{-2.73}$ & $51.39^{+11.57}_{-11.62}$ \\
		\hline 
		Depleted Mantle & $3.58^{+1.04}_{-1.02}$ & $0.85^{+0.50}_{-0.48}$ & $4.43^{+1.54}_{-1.50}$ \\
		Enriched Mantle & $4.23^{+1.84}_{-1.82}$ & $0.95^{+0.61}_{-0.60}$ & $5.18^{+2.43}_{-2.38}$ \\
		\hline 
		Total BSE & $47.20^{+8.68}_{-8.69}$ & $13.79^{+2.53}_{-2.54}$ & $60.99^{+10.91}_{-11.00}$ \\
		\bottomrule
	\end{tabular}
\end{table}

\begin{table}[t]
	\centering
	\caption{Geoneutrino flux at the Jinping detector predicted with the ECM1 model.}
	\label{tab:flux_jinping_ecm1}
	\renewcommand{\arraystretch}{1.4}
	\begin{tabular}{lccc}
		\toprule
		& \multicolumn{3}{c}{Flux (TNU)} \\
		& $U$ & $Th$ & $U + Th$ \\
		\midrule
		Sediment & $0.72^{+0.09}_{-0.09}$ & $0.23^{+0.03}_{-0.03}$ & $0.95^{+0.12}_{-0.12}$ \\
		Upper Crust & $24.06^{+6.18}_{-6.24}$ & $6.22^{+0.99}_{-1.01}$ & $30.29^{+7.06}_{-7.18}$ \\
		Middle Crust & $5.98^{+3.53}_{-3.49}$ & $2.21^{+1.65}_{-1.61}$ & $8.19^{+5.10}_{-5.06}$ \\
		Lower Crust & $1.66^{+1.11}_{-1.11}$ & $1.04^{+0.82}_{-0.89}$ & $2.70^{+1.95}_{-1.98}$ \\
		\hline 
		Crust & $32.42^{+7.53}_{-7.53}$ & $9.71^{+2.38}_{-2.34}$ & $42.12^{+9.62}_{-9.54}$ \\
		Lithospheric Mantle & $1.03^{+0.69}_{-0.74}$ & $0.34^{+0.24}_{-0.27}$ & $1.37^{+0.93}_{-1.00}$ \\
		\hline 
		Lithosphere & $33.44^{+7.70}_{-7.68}$ & $10.05^{+2.40}_{-2.42}$ & $43.49^{+9.76}_{-9.77}$ \\
		\hline 
		Depleted Mantle & $3.73^{+1.08}_{-1.07}$ & $0.89^{+0.52}_{-0.50}$ & $4.61^{+1.60}_{-1.57}$ \\
		Enriched Mantle & $4.28^{+1.86}_{-1.80}$ & $0.97^{+0.61}_{-0.60}$ & $5.24^{+2.41}_{-2.34}$ \\
		\hline 
		Total BSE & $41.44^{+7.15}_{-7.23}$ & $11.90^{+2.22}_{-2.25}$ & $53.35^{+9.13}_{-9.25}$ \\
		\bottomrule
	\end{tabular}
\end{table}

\begin{table}[t]
	\centering
	\caption{Geoneutrino flux at the Jinping detector predicted with the CRUST1.0 model.}
	\label{tab:flux_jinping_crust1_0}
	\renewcommand{\arraystretch}{1.4}
	\begin{tabular}{lccc}
		\toprule
		& \multicolumn{3}{c}{Flux (TNU)} \\
		& $U$ & $Th$ & $U + Th$ \\
		\midrule
		Sediment & $0.69^{+0.09}_{-0.09}$ & $0.22^{+0.03}_{-0.03}$ & $0.91^{+0.12}_{-0.12}$ \\
		Upper Crust & $28.34^{+7.21}_{-7.33}$ & $7.35^{+1.17}_{-1.18}$ & $35.69^{+8.33}_{-8.45}$ \\
		Middle Crust & $6.89^{+3.90}_{-3.74}$ & $2.77^{+1.83}_{-1.79}$ & $9.66^{+5.70}_{-5.51}$ \\
		Lower Crust & $0.97^{+0.61}_{-0.64}$ & $0.57^{+0.40}_{-0.47}$ & $1.54^{+1.02}_{-1.10}$ \\
		\hline 
		Crust & $36.89^{+8.65}_{-8.48}$ & $10.91^{+2.41}_{-2.39}$ & $47.80^{+10.68}_{-10.65}$ \\
		Lithospheric Mantle & $1.68^{+1.32}_{-1.37}$ & $0.56^{+0.45}_{-0.47}$ & $2.24^{+1.75}_{-1.84}$ \\
		\hline 
		Lithosphere & $38.58^{+9.11}_{-9.01}$ & $11.46^{+2.66}_{-2.62}$ & $50.04^{+11.41}_{-11.32}$ \\
		\hline 
		Depleted Mantle & $3.58^{+1.04}_{-1.02}$ & $0.85^{+0.50}_{-0.48}$ & $4.42^{+1.54}_{-1.50}$ \\
		Enriched Mantle & $4.23^{+1.84}_{-1.83}$ & $0.95^{+0.61}_{-0.60}$ & $5.18^{+2.43}_{-2.36}$ \\
		\hline 
		Total BSE & $46.38^{+8.52}_{-8.57}$ & $13.27^{+2.42}_{-2.44}$ & $59.64^{+10.64}_{-10.76}$ \\
		\bottomrule
	\end{tabular}
\end{table}

%\setstcolor{blue}
%\st{%
	The CRUST1.0 prediction presented in Table~\ref{tab:flux_jinping_crust1_0} is consistent with earlier estimates for the JNE site based on global crustal models. Sramek \textit{et al.}~\mbox{\cite{Sramek:2016}} obtained \mbox{$58.5^{+7.4}_{-7.2}$}~TNU using CRUST1.0, Wan \textit{et al.}~\mbox{\cite{Wan:2017}} reported \mbox{$59.4 \pm 7.6$}~TNU, and Wipperfurth \textit{et al.}~\mbox{\cite{Wipperfurth2020}} predicted \mbox{$59.9^{+12.2}_{-10.1}$}~TNU using CRUST1.0. To facilitate comparison with the geometric mean reported by Wipperfurth \textit{et al.}, we also express our CRUST1.0-only prediction using this convention, obtaining \mbox{$58.64^{+11.92}_{-9.91}$}~TNU, in close agreement with earlier estimates.%
%}

%{\color{blue}
	The Combined Model predicts a total geoneutrino signal of $60.99^{+10.91}_{-11.00}$~TNU at JNE, of which 34.42~TNU originates from the 1,957 grid cells comprising the Near-Field Model. This total is 1.35~TNU ($\sim2\%$) higher than the CRUST1.0-only prediction of 59.64~TNU. Within the replaced region, the Near-Field Model has lower \(V_p\) values and approximately 5\% less crustal mass than CRUST1.0. A reduction in mass would decrease the predicted signal if composition and spatial distribution were unchanged. However, the refined model assigns more mass to regions closer to the detector and less to more distant regions, increasing the distance-weighted contribution. In addition, the adopted \(V_p\)--composition relation for the middle and lower crust (Eqs.~\eqref{eq:Vp}--\eqref{eq:abundance}) associates lower \(V_p\) values with larger felsic fractions and higher HPE abundances. In our calculation, this increases the middle- and lower-crustal U contribution. Together, these effects produce a modest net increase in the predicted signal. Although small compared with the prediction uncertainties, this shift illustrates the influence of regional crustal structure and associated geochemical abundances.

    Across the three models, the upper crust contributes approximately 70\% of the lithospheric signal and nearly 60\% of the total signal, reflecting its high HPE abundances and the greater contribution per unit source mass from material close to the detector. Since all three models adopt the same upper-crustal HPE abundances, differences in this contribution arise from variations in upper-crustal thickness, density, and spatial mass distribution. Compared with CRUST1.0, ECM1 predicts a total signal at JNE that is 6.30~TNU ($\sim$10\%) lower, with approximately 90\% of this decrease originating in the upper crust. ECM1 divides each continental crystalline crustal column into fixed upper-, middle-, and lower-crustal fractions of 33\%, 33\%, and 34\%, respectively~\cite{ECM1:2023}. This partitioning reduces the upper-crustal thickness beneath the site from 23.6 to 17.5~km, while the total crustal thickness and total crustal U mass each differ by only approximately 1\%. Given the higher U abundance in the upper crust (2.7~ppm) than in the middle and lower crust (0.1--1.4~ppm), this redistribution of mass among the crustal layers accounts for most of the decrease in the predicted signal. 

%}

We quantify the relationship between the geoneutrino flux and radiogenic heat production. Fig.~\ref{fig:mc_signal} shows $7\times10^4$ Monte Carlo simulations of the total BSE geoneutrino flux at Jinping as a function of total BSE radiogenic heat production, calculated using the Combined Model. In these simulations, the BSE U abundance is sampled from a uniform distribution spanning 6--40~ppb, rather than fixed at the 20.3~ppb adopted in Sec.~\ref{sec:geochem}. This range encompasses the low-, medium-, and high-$Q$ BSE models, allowing the simulated flux--heat relationship to be used to translate a measured geoneutrino flux into a model-dependent constraint on Earth's radiogenic power. The color scale indicates the sample density. The solid black line and dashed lines show the mean predicted signal and its $\pm1\sigma$ bounds, respectively, at each value of radiogenic heat production. The vertical bands denote the radiogenic heat ranges associated with the low-, medium-, and high-$Q$ BSE models. 

%\begin{table}[t]
%	\centering
%	\caption{Uncertainty budget for the predicted geoneutrino flux.}
%	\label{tab:uncertainty_budget}
%	\begin{tabular}{l c c}% NOTE: avoid p{} col-spec: incompatible with revtex4-2 on TeX Live 2025 (array v2.6)
%		\toprule
%		\textbf{Source} & \textbf{Input parameters} & \textbf{Uncertainty} \\
%		\midrule
%		Geophysical & \parbox[t]{200pt}{$V_p$ ($\pm5\%$), Thickness ($\pm12\%$), Density ($\pm5\%$)} & $\sim 30\%$ \\
%		Geochemical & \parbox[t]{200pt}{Layer abundances, End-member compositions, and HPE Ratios} & $\sim 70\%$ \\
%		\bottomrule
%	\end{tabular}
%\end{table}

\subsection{Reactor Antineutrino Background} 
\begin{figure}[t]
	\centering
	\includegraphics[width=0.45\textwidth]{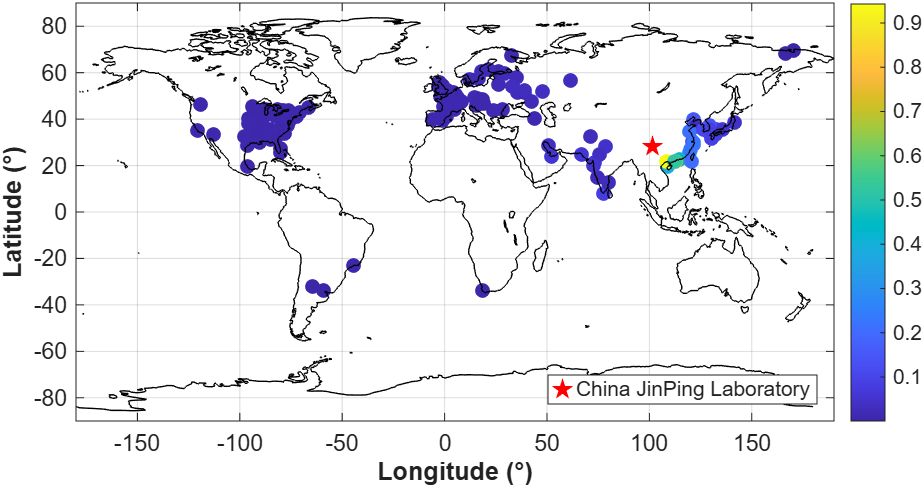}
	\caption{Worldwide distribution of the nuclear reactors considered in this study. The color of each marker indicates the predicted antineutrino signal contribution, in TNU.}
	\label{fig:WorldD}
\end{figure}

\begin{figure}[t]
	\centering
	\includegraphics[width=0.45\textwidth]{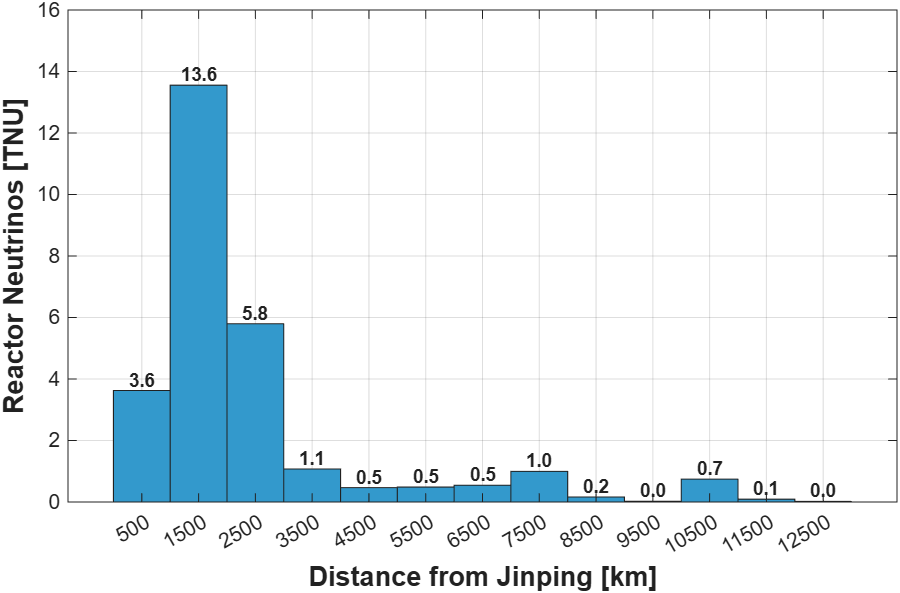}
	\caption{Predicted reactor antineutrino signal contributions, in TNU, at Jinping as a function of reactor--detector distance.}
	\label{fig:ReactorD}
\end{figure}

The global distribution of reactors considered in this study and their contributions to the antineutrino signal at Jinping are shown in Figs.~\ref{fig:WorldD} and~\ref{fig:ReactorD}. Fig.~\ref{fig:WorldD} maps the reactor locations, with marker colors indicating each reactor’s predicted signal contribution at Jinping, while Fig.~\ref{fig:ReactorD} presents these contributions as a function of reactor--detector distance. In the adopted reactor dataset, no nuclear power plant lies within approximately 950~km of the detector, and the dominant contributions come from China’s coastal reactor clusters at distances of $\gtrsim10^3$~km. The total predicted reactor antineutrino signal is 27.59~TNU, corresponding to approximately 41 IBD events for an exposure of 500~t $\times$ 1500~days as listed in Table~\ref{tab:CanRate}.

%\begin{figure*}[t]
%    \centering
%    \begin{subfigure}[b]{0.45\linewidth}
%        \centering
%        \includegraphics[width=\linewidth]{Fig1_Reactor_Distribution.png}
%        \label{fig:reactor_lonlat}
%    \end{subfigure}
%    \hfill
%    \begin{subfigure}[b]{0.45\linewidth}
%        \centering
%        \includegraphics[width=\linewidth]{Fig4_Distance_Distribution.png}
%        \label{fig:reactor_distance}
%    \end{subfigure}
%    \caption{Reactor neutrino contribution at Jinping: (a) worldwide distribution of the nuclear reactors considered in this study, where the color of each marker encodes the predicted antineutrino flux contribution (in TNU) of the corresponding reactor at the Jinping site; (b) reactor contributions as a function of the distance from Jinping.}
%    \label{fig:reactor_distribution}
%\end{figure*}

%The energy spectrum of the reactor neutrino background is presented in Fig.~\ref{Reactor_Energy_Spectrum}.
%\begin{figure}
 %   \centering
 %   \includegraphics[width=0.5\linewidth]{fig_reactor_bkg.png}
 %   \caption{The energy spectrum of the reactor neutrino background.}
 %   \label{Reactor_Energy_Spectrum}
%\end{figure}
The reactor antineutrino spectrum extends from the IBD threshold to approximately 8~MeV, well beyond the geoneutrino energy window. Below the $^{238}$U endpoint of 3.27~MeV, reactor antineutrinos overlap with the geoneutrino signal and constitute a major background to the measurement. Above approximately 3.3~MeV, the geoneutrino contribution vanishes and reactor antineutrinos dominate the IBD sample, allowing the reactor flux normalization to be constrained directly from the data.

\subsection{Combined IBD Spectrum and Event Rates}
\begin{figure}[t]
	\centering
	\includegraphics[width=0.45\textwidth]{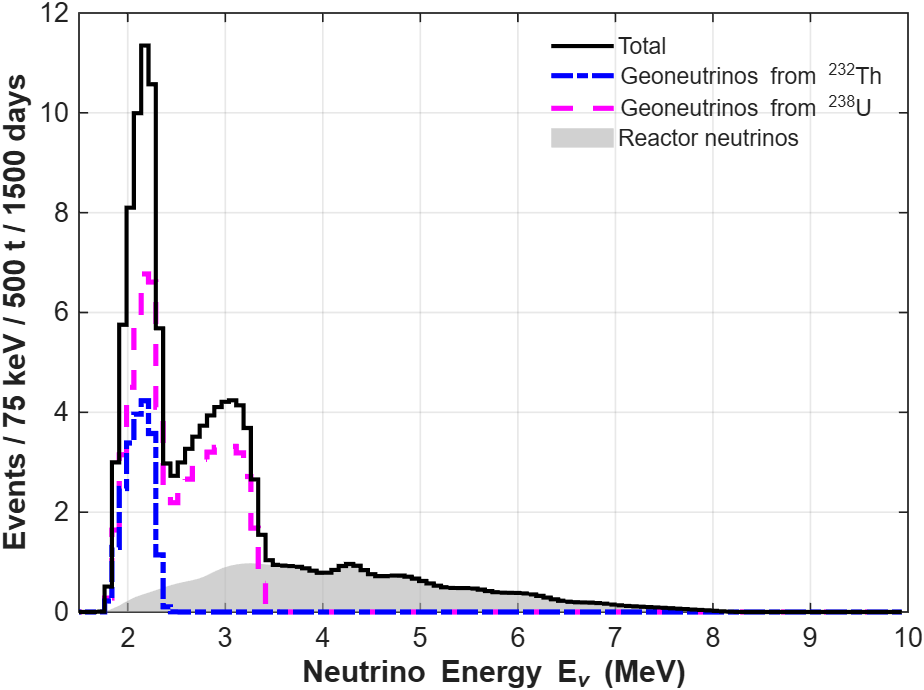}
	\caption{Predicted IBD event spectrum at the Jinping site, showing contributions from the $^{238}$U decay chain (magenta dashed), the $^{232}$Th decay chain (blue dash-dotted), and reactor antineutrinos (gray shaded). The solid black line represents the total spectrum.}
	\label{fig:eevent}
\end{figure}

This section presents the expected IBD spectrum and event rates at JNE by combining the predicted geoneutrino signal with the reactor antineutrino background, providing a quantitative reference for future data analysis. The geoneutrino calculation uses the Combined Model, and the adopted rates are the arithmetic means of the Monte Carlo simulations obtained in Sec.~\ref{sec:results_flux} and listed in Table~\ref{tab:flux_jinping_crust1_0_swchinacvm}. Fig.~\ref{fig:eevent} shows the expected IBD spectrum for an exposure of 500~t $\times$ 1500~days, assuming an energy resolution of $4.4\%/\sqrt{E_{\mathrm{vis}}/\mathrm{MeV}}$, corresponding to a light yield of 500~p.e./MeV. The spectrum includes both geoneutrino and reactor antineutrino contributions. 

The detectable geoneutrino energy window extends from the IBD threshold of 1.806~MeV to the $^{238}$U spectrum endpoint of 3.27~MeV, while the $^{232}$Th spectrum ends at 2.25~MeV. The two components therefore overlap only below 2.25~MeV, with the $^{238}$U chain alone contributing to the geoneutrino signal between the two endpoints. These boundaries refer to the true antineutrino energy. Finite detector energy resolution smears the corresponding features in the reconstructed energy spectrum, as shown in Fig.~\ref{fig:eevent}. The distinct spectral shapes allow the U and Th contributions to be separated through a spectral fit. The reactor antineutrino spectrum spans the entire geoneutrino energy window and extends to higher energies. The predicted geoneutrino and reactor antineutrino event rates, expressed in TNU, and the corresponding event counts for an exposure of 500~t $\times$ 1500~days are summarized in Table~\ref{tab:CanRate}.
\begin{table}[!h]
	\centering
	\caption{Predicted geoneutrino and reactor antineutrino event rates of full spectrum in TNU and corresponding event counts for an exposure of 500~t $\times$ 1500~days.}
	\label{tab:CanRate}
	\begin{tabular}{llcclr} \hline\hline
		& \multicolumn{3}{c}{Geoneutrino} & \multicolumn{2}{c}{Reactor $\bar{\nu}_e$} \\
		& $^{238}$U & $^{232}$Th & $^{238}$U + $^{232}$Th &   \\\hline
		Event Rates (TNU)   & 47.20 & 13.79 & 60.99 & 27.59  \\
		Event Counts       & 69.78 & 20.39 & 90.18 & 40.78  \\\hline\hline
	\end{tabular}
\end{table}
The $^{238}$U chain provides the largest contribution to the predicted geoneutrino signal, while the crust dominates both the total signal and its systematic uncertainty. Improved constraints on the local crustal structure and composition will therefore complement increased exposure, particularly in extracting the mantle signal, consistent with earlier studies highlighting the importance of refined local crustal models~\cite{Sramek:2016, Wipperfurth2020}.

%============================================================
\section{Summary}
\label{sec:summary}
%============================================================
We have presented a refined prediction of the geoneutrino signal at Jinping for the 500-tonne JNE, currently under construction at CJPL. To better characterize the local crust, which contributes substantially to the signal, we constructed the Near-Field Model using the SWChinaCVM-2.0 regional velocity model and Moho depths from SWChinaCVM-MOHO-2.0. The model refines the crustal layer geometry, seismic velocities, densities and HPE abundances on a $0.2^\circ \times 0.2^\circ$ grid and is integrated with the global CRUST1.0 model.

Assuming the medium-$Q$ BSE composition, the Combined Model predicts a total geoneutrino signal of $60.99^{+10.91}_{-11.00}$~TNU, including a lithospheric contribution of $51.39^{+11.57}_{-11.62}$~TNU. Incorporating the Near-Field Model increases the predicted signal by 1.35~TNU ($\sim$2\%) relative to CRUST1.0 alone, while the ECM1 prediction is 6.30~TNU ($\sim$ 10\%) lower than the CRUST1.0 prediction. The difference between the two global predictions exceeds the shift introduced by the near-field refinement. Together, these comparisons highlight the importance of both the global crustal model and local refinements in predicting the geoneutrino signal at Jinping.

The predicted reactor antineutrino background integrated over the full spectrum above the IBD threshold is 27.59~TNU, less than half the geoneutrino signal, with only approximately 25\% falling within the geoneutrino energy window. This relatively low background reflects Jinping’s remoteness from nuclear power plants, with the nearest plant in the adopted reactor dataset located approximately 950~km away. 

For an exposure of 500~t $\times$ 1500~days, approximately 90 geoneutrino and 41 reactor IBD events are expected. The distinct spectral shapes of the $^{238}$U, $^{232}$Th, and reactor components provide a basis for their statistical separation. The reactor-dominated region above a neutrino energy of 3.3~MeV provides an in situ constraint on the reactor flux normalization, supporting the extraction of the total geoneutrino signal. The mantle contribution can then be inferred by subtracting the predicted lithospheric contribution. These predictions provide a quantitative reference for future JNE analyses, particularly for constraining the mantle signal and testing competing BSE models. 

\section{ACKNOWLEDGMENTS}
		We thank Dr. Laura S. Keen for sharing the GEONU calculation framework and Dr. Weilai Wang for providing the SWChinaCVM-MOHO-2.0 Moho interface model. This work is supported by the Discipline Construction Fund of Shandong University. 

\bibliography{jne_refs}

@incollection{Jaupart:2007,
  author    = {Jaupart, C. and Labrosse, S. and Mareschal, J.-C.},
  title     = {Temperatures, heat and energy in the mantle of the {E}arth},
  booktitle = {Treatise on Geophysics},
  editor    = {Schubert, G.},
  publisher = {Elsevier},
  pages     = {253--303},
  year      = {2007},
  doi       = {10.1016/B978-044452748-6.00114-0}
}

@article{JH:2010,
  author  = {Davies, J. H. and Davies, D. R.},
  title   = {{E}arth's surface heat flux},
  journal = {Solid Earth},
  volume  = {1},
  pages   = {5},
  year    = {2010},
  doi     = {10.5194/se-1-5-2010}
}

@article{Javoy:2010,
  author  = {Javoy, M. and others},
  title   = {The chemical composition of the {E}arth: {E}nstatite chondrite models},
  journal = {Earth Planet. Sci. Lett.},
  volume  = {293},
  pages   = {259},
  year    = {2010},
  doi     = {10.1016/j.epsl.2010.02.033}
}

@article{ArevaloJr:2010,
  author  = {Arevalo Jr., R. and McDonough, W. F.},
  title   = {Chemical variations and regional diversity observed in {MORB}},
  journal = {Chem. Geol.},
  volume  = {271},
  pages   = {70},
  year    = {2010},
  doi     = {10.1016/j.chemgeo.2009.12.013}
}

@article{Arevalo2009,
  author  = {Arevalo Jr., R. and McDonough, W. F. and Luong, M.},
  title   = {The {K/U} ratio of the silicate {E}arth: {I}nsights into mantle composition, structure and thermal evolution},
  journal = {Earth Planet. Sci. Lett.},
  volume  = {278},
  pages   = {361},
  year    = {2009},
  doi     = {10.1016/j.epsl.2008.12.023}
}

@article{Arevalo2013,
  author  = {Arevalo Jr., R. and McDonough, W. F. and Stracke, A. and Willbold, M. and Ireland, T. J. and Walker, R. J.},
  title   = {Simplified mantle architecture and distribution of radiogenic power},
  journal = {Geochem. Geophys. Geosyst.},
  volume  = {14},
  pages   = {2265},
  year    = {2013},
  doi     = {10.1002/ggge.20152}
}

@article{Wipperfurth2018,
  author  = {Wipperfurth, S. A. and Guo, M. and {\v{S}}r{\'a}mek, O. and McDonough, W. F.},
  title   = {Earth's chondritic {Th/U}: {N}egligible fractionation during accretion, core formation, and crust--mantle differentiation},
  journal = {Earth Planet. Sci. Lett.},
  volume  = {498},
  pages   = {196},
  year    = {2018},
  doi     = {10.1016/j.epsl.2018.06.029}
}

@article{turcotte:1984,
  author  = {Turcotte, D. L. and Schubert, G.},
  journal = {Geol. J.},
  volume  = {19},
  pages   = {303},
  year    = {1984}
}

@book{Deutsch:1998,
  author    = {Deutsch, C. V. and Journel, A. G.},
  title     = {{GSLIB}: Geostatistical Software Library and User's Guide},
  edition   = {2nd},
  publisher = {Oxford University Press},
  address   = {New York},
  year      = {1998}
}

@article{McDonough:2014,
  author  = {McDonough, W. F. and Sramek, O.},
  title   = {Neutrino geoscience, news in brief},
  journal = {Environ. Earth Sci.},
  volume  = {71},
  pages   = {3787},
  year    = {2014},
  doi     = {10.1007/s12665-014-3133-9}
}

@article{Dye:2012,
  author  = {Dye, S. T.},
  title   = {Geoneutrinos and the radioactive power of the {E}arth},
  journal = {Rev. Geophys.},
  volume  = {50},
  pages   = {RG3007},
  year    = {2012},
  doi     = {10.1029/2012RG000400}
}

@article{Zhu:2026,
  author  = {Zhu, G. and Huang, H.},
  title   = {Ultradeep drilling beyond 10 km revealing new insights into {E}arth systems and resources},
  journal = {Commun. Earth Environ.},
  volume  = {7},
  pages   = {124},
  year    = {2026},
  doi     = {10.1038/s43247-026-03246-z}
}

@article{KamLAND-2005,
  author  = {Araki, T. and others},
  title   = {Experimental investigation of geologically produced antineutrinos with {K}am{LAND}},
  journal = {Nature},
  volume  = {436},
  pages   = {499--503},
  year    = {2005},
  doi     = {10.1038/nature03980}
}

@article{KamLAND-2011,
  author  = {Gando, A. and others},
  title   = {Partial radiogenic heat model for {E}arth revealed by geoneutrino measurements},
  journal = {Nat. Geosci.},
  volume  = {4},
  pages   = {647--651},
  year    = {2011},
  doi     = {10.1038/ngeo1205}
}

@article{KamLAND-2013,
  author        = {Gando, A. and others},
  collaboration = {KamLAND},
  title         = {Reactor on-off antineutrino measurement with {K}am{LAND}},
  journal       = {Phys. Rev. D},
  volume        = {88},
  pages         = {033001},
  year          = {2013},
  doi           = {10.1103/PhysRevD.88.033001}
}

@article{KamLAND-2022,
  author        = {Abe, S. and others},
  collaboration = {KamLAND},
  title         = {Abundances of uranium and thorium elements in {E}arth estimated by geoneutrino spectroscopy},
  journal       = {Geophys. Res. Lett.},
  volume        = {49},
  pages         = {e2022GL099566},
  year          = {2022},
  doi           = {10.1029/2022GL099566}
}

@article{Borexino-2010,
  author        = {Bellini, G. and others},
  collaboration = {Borexino},
  title         = {Observation of geo-neutrinos},
  journal       = {Phys. Lett. B},
  volume        = {687},
  pages         = {299--304},
  year          = {2010},
  doi           = {10.1016/j.physletb.2010.03.051}
}

@article{Borexino-2013,
  author        = {Bellini, G. and others},
  collaboration = {Borexino},
  title         = {Measurement of geo-neutrinos from 1353 days of {B}orexino},
  journal       = {Phys. Lett. B},
  volume        = {722},
  pages         = {295--300},
  year          = {2013},
  doi           = {10.1016/j.physletb.2013.04.030}
}

@article{Borexino-2020,
  author        = {Agostini, M. and others},
  collaboration = {Borexino},
  title         = {Comprehensive geoneutrino analysis with {B}orexino},
  journal       = {Phys. Rev. D},
  volume        = {101},
  pages         = {012009},
  year          = {2020},
  doi           = {10.1103/PhysRevD.101.012009}
}

@article{SNO+-2025-1,
  author        = {Abreu, M. and others},
  collaboration = {SNO+},
  title         = {Measurement of reactor antineutrino oscillation at {SNO+}},
  journal       = {Phys. Rev. Lett.},
  volume        = {135},
  pages         = {121801},
  year          = {2025},
  doi           = {10.1103/gypt-lc9v}
}

@article{SNO:2025chx,
  author        = {Abreu, M. and others},
  collaboration = {SNO+},
  title         = {Measurement of reactor antineutrino oscillations with 1.46 kilotonne-years of data at {SNO+}},
  journal       = {Phys. Rev. Lett.},
  volume        = {137},
  pages         = {101807},
  year          = {2026},
  doi           = {10.1103/mv5z-pfb7}
}

@article{JUNO:2025gmd,
  author        = {Abusleme, A. and others},
  collaboration = {JUNO},
  title         = {Measurement of reactor neutrino oscillation with the first {JUNO} data},
  journal       = {Nature},
  volume        = {654},
  pages         = {343--348},
  year          = {2026},
  doi           = {10.1038/s41586-026-10538-z}
}

@article{JinpingNE:2016,
  author        = {Beacom, J. F. and others},
  collaboration = {Jinping},
  title         = {Physics prospects of the {J}inping neutrino experiment},
  journal       = {Chin. Phys. C},
  volume        = {41},
  pages         = {023002},
  year          = {2017},
  doi           = {10.1088/1674-1137/41/2/023002}
}

@article{Sramek:2016,
  author  = {Sramek, O. and others},
  title   = {Revealing the {E}arth's mantle from the tallest mountains using the {J}inping {N}eutrino {E}xperiment},
  journal = {Sci. Rep.},
  volume  = {6},
  pages   = {33034},
  year    = {2016},
  doi     = {10.1038/srep33034}
}

@article{Wan:2017,
  author  = {Wan, L. and Hussain, G. and Wang, Z. and Chen, S.},
  title   = {Geoneutrinos at {J}inping: {F}lux prediction and oscillation analysis},
  journal = {Phys. Rev. D},
  volume  = {95},
  pages   = {053001},
  year    = {2017},
  doi     = {10.1103/PhysRevD.95.053001}
}

@article{PDG:2025,
  author        = {Navas, S. and others},
  collaboration = {Particle Data Group},
  title         = {Review of particle physics},
  journal       = {Phys. Rev. D},
  volume        = {110},
  pages         = {030001},
  year          = {2024},
  doi           = {10.1103/PhysRevD.110.030001},
  note          = {2025 update}
}

@article{Enomoto:2005,
  author  = {Enomoto, S. and Ohtani, E. and Inoue, K. and Suzuki, A.},
  title   = {Neutrino geophysics with {K}am{LAND} and future prospects},
  journal = {Earth Planet. Sci. Lett.},
  volume  = {258},
  pages   = {147--159},
  year    = {2007},
  doi     = {10.1016/j.epsl.2007.03.046}
}

@article{CRUST:2014,
  author  = {Laske, G. and others},
  title   = {Update on {CRUST1.0} - {A} 1-degree global model of {E}arth's crust},
  journal = {Geophys. Res. Abstracts},
  volume  = {15},
  pages   = {EGU2013},
  year    = {2013}
}

@article{LITHO1.0-2014,
  author  = {Pasyanos, M. E. and Masters, T. G. and Laske, G. and Ma, Z.},
  title   = {{LITHO1.0}: An updated crust and lithospheric model of the {E}arth},
  journal = {J. Geophys. Res. Solid Earth},
  volume  = {119},
  pages   = {2153--2173},
  year    = {2014},
  doi     = {10.1002/2013JB010626}
}

@article{Huang:2013,
  author  = {Huang, Y. and others},
  title   = {A reference {E}arth model for the heat-producing elements and associated geoneutrino flux},
  journal = {Geochem. Geophys. Geosyst.},
  volume  = {14},
  pages   = {2003},
  year    = {2013},
  doi     = {10.1002/ggge.20129}
}

@article{Vogel:1999,
  author  = {Vogel, P. and Beacom, J. F.},
  title   = {Angular distribution of neutron inverse beta decay, {$\bar{\nu}_{e}+p\rightarrow e^{+}+n$}},
  journal = {Phys. Rev. D},
  volume  = {60},
  pages   = {053003},
  year    = {1999},
  doi     = {10.1103/PhysRevD.60.053003}
}

@article{Huber:2011,
  author  = {Huber, P.},
  title   = {Determination of antineutrino spectra from nuclear reactors},
  journal = {Phys. Rev. C},
  volume  = {84},
  pages   = {024617},
  year    = {2011},
  doi     = {10.1103/PhysRevC.84.024617}
}

@article{Mueller:2011,
  author  = {Mueller, T. A. and others},
  title   = {Improved predictions of reactor antineutrino spectra},
  journal = {Phys. Rev. C},
  volume  = {83},
  pages   = {054615},
  year    = {2011},
  doi     = {10.1103/PhysRevC.83.054615}
}

@misc{IAEA:2015,
  author       = {{International Atomic Energy Agency}},
  title        = {Power Reactor Information System ({PRIS})},
  howpublished = {\url{https://pris.iaea.org/}},
  year         = {2015}
}

@misc{INFN:antineutrino,
  author       = {{Istituto Nazionale di Fisica Nucleare}},
  title        = {A Reference Worldwide Model for Antineutrinos from Reactors},
  howpublished = {\url{https://www.fe.infn.it/radioactivity/antineutrino/index.html}},
  year         = {2015}
}

@article{Baldoncini:2015,
  author  = {Baldoncini, M. and others},
  title   = {Reference worldwide model for antineutrinos from reactors},
  journal = {Phys. Rev. D},
  volume  = {91},
  pages   = {065002},
  year    = {2015},
  doi     = {10.1103/PhysRevD.91.065002}
}

@article{Ma:2012,
  author  = {Ma, X. B. and others},
  title   = {Improved calculation of the energy release in neutron-induced fission},
  journal = {Phys. Rev. C},
  volume  = {88},
  pages   = {014605},
  year    = {2013},
  doi     = {10.1103/PhysRevC.88.014605}
}

@article{Wipperfurth2020,
  author  = {Wipperfurth, S. A. and {\v{S}}r{\'a}mek, O. and McDonough, W. F.},
  title   = {Reference models for lithospheric geoneutrino signal},
  journal = {J. Geophys. Res. Solid Earth},
  volume  = {125},
  pages   = {e2019JB018433},
  year    = {2020},
  doi     = {10.1029/2019JB018433}
}

@article{Liu2023,
  author  = {Liu, Y. and Yu, Z. and Zhang, Z. and Yao, H. and Wang, W. and Zhang, H. and Fang, H. and Fang, L.},
  title   = {The high-resolution community velocity model {V2.0} of southwest {C}hina, constructed by joint body and surface wave tomography of data recorded at temporary dense arrays},
  journal = {Sci. China Earth Sci.},
  volume  = {66},
  pages   = {2368},
  year    = {2023},
  doi     = {10.1007/s11430-022-1161-7}
}

@article{Liu2025,
  author  = {Liu, H. J. and Wang, W. L. and Zhang, L. and Cai, G. Y. and Bao, J. J.},
  title   = {Community {M}oho interface model in {SE} {T}ibet based on several new generation velocity models},
  journal = {Chinese J. Geophys.},
  volume  = {68},
  pages   = {827},
  year    = {2025},
  doi     = {10.6038/cjg2024S0087}
}

@article{Wu2024,
  author  = {Wu, J. P. and Cai, Y. and Wang, W. and others},
  title   = {Three dimensional velocity model and its tectonic implications at {C}hina seismic experimental site, eastern margin of the {T}ibetan {P}lateau},
  journal = {Sci. China Earth Sci.},
  year    = {2024},
  doi     = {10.1007/s11430-023-1293-4}
}

@article{Han2022,
  author  = {Han, S. C. and Zhang, H. J. and Xin, H. L. and others},
  title   = {{USTClitho2.0}: Updated unified seismic tomography models for continental {C}hina lithosphere from joint inversion of body-wave arrival times and surface-wave dispersion data},
  journal = {Seismol. Res. Lett.},
  volume  = {93},
  pages   = {201},
  year    = {2022},
  doi     = {10.1785/0220210122}
}

@article{Sramek:2013,
  author  = {Sramek, O. and McDonough, W. F. and Kite, E. S. and Lekic, V. and Dye, S. T. and Zhong, S.},
  title   = {Geophysical and geochemical constraints on geoneutrino fluxes from {E}arth's mantle},
  journal = {Earth Planet. Sci. Lett.},
  volume  = {361},
  pages   = {356},
  year    = {2013},
  doi     = {10.1016/j.epsl.2012.11.001}
}

@article{Brocher:2005,
  author  = {Brocher, T. M.},
  title   = {Empirical relations between elastic wavespeeds and density in the {E}arth's crust},
  journal = {Bull. Seismol. Soc. Am.},
  volume  = {95},
  pages   = {2081},
  year    = {2005},
  doi     = {10.1785/0120050077}
}

@article{ECM1:2023,
  author  = {Mooney, W. D. and Barrera-Lopez, C. and Su{\'a}rez, M. G. and Castelblanco, M. A.},
  title   = {{E}arth {C}rustal {M}odel 1 ({ECM1}): {A} 1$^\circ$$\times$1$^\circ$ global seismic and density model},
  journal = {Earth-Sci. Rev.},
  volume  = {243},
  pages   = {104493},
  year    = {2023},
  doi     = {10.1016/j.earscirev.2023.104493}
}

@incollection{RudnickGao:2014,
  author    = {Rudnick, R. L. and Gao, S.},
  title     = {Composition of the continental crust},
  booktitle = {Treatise on Geochemistry},
  edition   = {2nd},
  volume    = {4},
  editor    = {Holland, H. D. and Turekian, K. K.},
  publisher = {Elsevier},
  pages     = {1--51},
  year      = {2014},
  doi       = {10.1016/B978-0-08-095975-7.00301-6}
}

@incollection{Plank:2014,
  author    = {Plank, T.},
  title     = {The chemical composition of subducting sediments},
  booktitle = {Treatise on Geochemistry},
  edition   = {2nd},
  volume    = {4},
  editor    = {Holland, H. D. and Turekian, K. K.},
  publisher = {Elsevier},
  pages     = {607--629},
  year      = {2014},
  doi       = {10.1016/B978-0-08-095975-7.00319-3}
}

@incollection{WhiteKlein:2014,
  author    = {White, W. M. and Klein, E. M.},
  title     = {Composition of the oceanic crust},
  booktitle = {Treatise on Geochemistry},
  edition   = {2nd},
  volume    = {4},
  editor    = {Holland, H. D. and Turekian, K. K.},
  publisher = {Elsevier},
  pages     = {457--496},
  year      = {2014},
  doi       = {10.1016/B978-0-08-095975-7.00315-6}
}

@article{McDonoughSun:1995,
  author  = {McDonough, W. F. and Sun, S.-s.},
  title   = {The composition of the {E}arth},
  journal = {Chem. Geol.},
  volume  = {120},
  pages   = {223},
  year    = {1995},
  doi     = {10.1016/0009-2541(94)00140-4}
}

@techreport{Amante2009,
  author      = {Amante, Christopher and Eakins, Barry W.},
  title       = {{ETOPO1} 1 Arc-Minute Global Relief Model: Procedures, Data Sources and Analysis},
  institution = {National Geophysical Data Center, NOAA},
  type        = {NOAA Technical Memorandum},
  number      = {NESDIS NGDC-24},
  year        = {2009},
  doi         = {10.7289/V5C8276M}
}

\end{document}